\documentclass[12pt]{article}
\usepackage[utf8]{inputenc}
\usepackage{amsmath,amssymb,amsfonts,amsthm}
\usepackage{graphicx}
\usepackage{bbm}
\usepackage[caption=false]{subfig}
\usepackage[pdftex,bookmarks=false,colorlinks=true,linkcolor=blue,
citecolor=blue,filecolor=black,urlcolor=blue]{hyperref}

\numberwithin{equation}{section}

\begin{document}
\begin{titlepage}
\begin{center}
\vspace*{2cm} {\Large {\bf  Equilibrium time-correlation functions for one-dimensional hard-point systems \bigskip\bigskip\\}}
{\large Christian B.~Mendl$^1$ and Herbert Spohn$^2$}\bigskip\bigskip\\
$^1$Zentrum Mathematik, TU M\"unchen, Boltzmannstra{\ss}e 3, D-85747 Garching, Germany\\
$^2$Institute for Advanced Study, Einstein Drive, Princeton New Jersey 08540, USA\\  Zentrum Mathematik and Physik Department, TU M\"unchen,\\
Boltzmannstra{\ss}e 3, D-85747 Garching, Germany\\
e-mail:~{\tt mendl@ma.tum.de}, {\tt spohn@ma.tum.de}\\
\end{center}
\vspace{5cm} \textbf{Abstract.} As recently proposed, the long-time behavior of equilibrium time-correlation functions for one-dimensional systems are expected to be captured by a nonlinear extension of fluctuating hydrodynamics. We outline the 
predictions from the theory aimed at the  comparison with molecular dynamics. We report on numerical simulations of a fluid with a hard-shoulder potential and of a hard-point gas with alternating masses. These models have in common that the collision time is zero and their dynamics amounts to iterating collision by collision. The theory is well confirmed, with the twist that the non-universal coefficients are still changing at longest accessible times.

\end{titlepage}

\section{Introduction}
\label{sec1}

As very well understood, in thermal equilibrium one-dimensional classical fluids  show no phase transitions and have rapidly decaying static correlations, provided the interaction potential is sufficiently short-ranged \cite{Ru69}. On the other hand, as discovered in the early 1970ies, time correlations have anomalous decay. In particular, total current-current correlations decay non-integrably and the Green-Kubo definition of transport coefficients yields divergent expressions \cite{AlWa70,ErHa76}. At the time only fairly qualitative predictions were available. But over the last 15 years there has been  a considerable spectrum of molecular dynamics (MD) simulations,
which do provide quantitative information \cite{LeLi03,Dh08}.  Currently the conventional system size is of the order of $10^4$ particles and the 
maximal simulation time is such that the right and left going sound modes  first collide in a ring geometry. Most efforts have been directed towards the numerical value of the dynamical exponents and the issue of universality. Very recently, in addition to exponents, universal scaling functions have been proposed on the basis of nonlinear fluctuating hydrodynamics \cite{vB12,MS13}. The main goal of our contribution is to compare these theoretical predictions with MD simulations.

For such purpose we consider hard-point systems, mainly because there is then no need to simulate differential equations.
The dynamics proceeds from collision to collision with free motion in-between. Such models have been studied extensively before \cite{GrNa02,DeNa03,CiDe05,DeDe07,Po11,ZaDe13,Ge12,vB13}. While there are important hints in the literature, the available simulations are not specific enough to test the theory. Therefore we decided to redo three of the most common models. (1) a hard-point gas of particles with equal mass and interaction between neighbors through a ``shoulder potential", (2) a hard-point gas of particles with alternating masses, and (3) the same as (2) but
with the hard-point potential replaced by an infinitely high square-well potential. In other words, when neighboring particles
reach a maximal distance, say $a$, then there is an inward collision.

These models have in common that they are particular instances of anharmonic chains, as characterized by having  an interaction only
between particles of adjacent label. For hard-point systems, in addition, the spatial order coincides with the label order.
The equilibrium measure of a generic anharmonic chain is of product form. Therefore no equilibration step is required. The true equilibrium distribution is swiftly produced by a random number generator. We believe that this is of advantage as compared to the more conventional dynamical equilibration, which has always the risk of systematic errors, even though the
necessarily limited numerical tests indicate an equilibrated system.

Nonlinear fluctuating hydrodynamics makes the implicit assumptions that there are no further local conservation laws beyond the three standard ones  and that upon fixing their values the dynamics is sufficiently well mixing in time. These assumptions cannot be checked easily.  Counterexamples are completely integrable chains, as the Toda chain.
For the shoulder potential the scattering induced through  the potential step seems to suffice. Models (2) and (3) from
above become integrable in case of equal masses. Presumably any other mass ratio destroys their integrability.
Based on the experience from MD simulations a mass ratio around 3 is sufficiently well mixing. 

As an outline: In the following two sections we introduce the hard-point systems under study and review the theory.
More details are recorded in \cite{Sp13}. In Section 4 we report our MD results and in Section 5 we arrive at conclusions and
compare with other MD data available.

\section{Hard-point systems}
\label{sec2}

\paragraph{Monoatomic chains.} The hamiltonian of a one-dimensional fluid is of the form
\begin{equation}\label{2.1}
H_\mathrm{f}=\sum^N_{j=1} \tfrac{1}{2}p^2_j +\tfrac{1}{2}\sum^N_{i\neq j=1}V(q_i-q_j)\,.
\end{equation}
Here $q_j$ is the position and $p_j$ the momentum of the $j$-th particle. Momentum equals velocity, since we use units for which the mass equals 1. $V$ is the interaction potential. We now choose specifically the \emph{shoulder potential}
\begin{equation}
\label{2.2}
V_\mathrm{sh}(x) = \infty \mathrm{\hspace{4pt}for\hspace{4pt}} |x|\leq \tfrac{1}{2}\,,\quad V_\mathrm{sh}(x) = 1 \mathrm{\hspace{4pt}for\hspace{4pt}} \tfrac{1}{2}< |x|  < 1\,,\quad
V_\mathrm{sh}(x) = 0 \mathrm{\hspace{4pt}for\hspace{4pt}} 1\leq |x|\,.
\end{equation}
If one initially imposes $q_j +\tfrac{1}{2}\leq q_{j + 1}$, then this order is preserved in time and the interaction is only between neighboring particles. We introduce the $j$-th stretch
\begin{equation}\label{2.3}
r_j = q_{j+1} - q_j \,.
\end{equation}
Then 
\begin{equation}\label{2.4}
\dot{r}_j = p_{j+1} - p_j\,,\quad \dot{p}_j = V'_\mathrm{sh}(r_j)- V'_\mathrm{sh}(r_{j-1})\,.
\end{equation}
Without loss of generality, the potential height is chosen to be $1$ and the hard core size to be $\tfrac{1}{2}$. The width of the potential step could be any value between $0$ and $\tfrac{1}{2}$. We study here only the maximal width.

It is of advantage to view $(r_j,p_j)_{j = 1,...,N}$ as a one-dimensional field theory with two components. Periodic boundary conditions, $r_{j+N} = r_j$, $p_{j+N} = p_j$, are imposed throughout. For hard-point particles $r_j\geq 0$ and $p_j \in \mathbb{R}$. The somewhat singular
force in \eqref{2.4} translates into the following collision rules. Between collisions one has free motion with $\dot{p}_j = 0$.
There are two types of collisions, at $r_j = \tfrac{1}{2}$ and at $r_j =  1$.\medskip\\ 
\textit{(i) $r_j = \frac{1}{2}$}.  If $p_{j+1} - p_j <0$,
then there is a point collision as
\begin{equation}\label{2.6}
\begin{split}
p_j' &= p_{j+1}\,, \\
p_{j+1}' &= p_{j}\,,
\end{split}
\end{equation}
where $'$ denotes the momentum after the collision.  If $p_{j+1} - p_j > 0$, particles separate under free motion.\medskip\\
\textit{(ii)} $r_j = 1$. There is scattering at the potential step depending on whether particles approach or recede from each other. In the latter case, i.e., $p_{j+1} - p_{j} > 0$, the collision rule reads
\begin{equation}\label{2.7}
\begin{split}
p_j' &= \tfrac{1}{2}\Big(p_j + p_{j+1} - \sqrt{(p_{j+1} - p_j)^2 + 4}\Big)\,, \\
p_{j+1}' &= \tfrac{1}{2}\Big(p_j + p_{j+1} + \sqrt{(p_{j+1} - p_j)^2 + 4}\Big) \,.
\end{split}
\end{equation}
For approaching particles with large momentum difference $p_{j} - p_{j+1} > 2$, the momentum transfer is sufficient to enter the shoulder plateau and the collision rule reads
\begin{equation}\label{2.8}
\begin{split}
p_j' &= \tfrac{1}{2}\Big(p_j + p_{j+1} + \sqrt{(p_{j+1} - p_j)^2 - 4}\Big)\,, \\
p_{j+1}' &= \tfrac{1}{2}\Big(p_j + p_{j+1} - \sqrt{(p_{j+1} - p_j)^2 - 4}\Big) \,.
\end{split}
\end{equation}
If the incoming momentum transfer is too small, then the particles are specularly reflected, i.e., if $0 < p_{j} - p_{j+1} < 2$, then
\begin{equation}\label{2.9}
\begin{split}
p_j' &=  p_{j+1}\,, \\
p_{j+1}' &= p_{j}\,.
\end{split}
\end{equation}

An anharmonic chain, in general,  still evolves according to \eqref{2.4}, but with $V_\mathrm{sh}$ replaced by some potential $V$ and generically without constraints on $r_j$.
$V$ is assumed to be bounded from below and to have at least a one-sided linearly increasing bound at infinity.
Thermal equilibrium is described by the canonical Gibbs measure at zero average momentum. It is given by a product measure, i.e., the $(r_j,p_j)_{j=1,...,N}$ are independent. At a single site, the momentum $p_j$ has a Maxwellian distribution with mean zero and variance $1/2\beta$,
while the probability density of the stretch $r_j$ is given by
\begin{equation}\label{2.10}
Z^{-1} \mathrm{e}^{-\beta(V(y)+py)} \,,\quad Z(p,\beta)=\int_\mathbb{R}dy \mathrm{e}^{-\beta(V(y)+py)}\,.
\end{equation}
$p$ controls the stretch and $\beta$ the energy. Averages with respect to~\eqref{2.10} are denoted by $\langle\cdot\rangle_{p,\beta}$. Note that
\begin{equation}\label{2.11}
  p=-\langle V'(y)\rangle_{p,\beta}
\end{equation}
and, as average force on a specified particle, $p$ is identified with the thermodynamic pressure.

In our simulations the average is always with respect to this canonical equilibrium measure. In the literature one finds an effort
to impose zero momentum strictly and not only on average. But for the sizes and time spans
under investigation, this makes hardly any difference. 

For the purpose of nonlinear fluctuating hydrodynamics we also record the Euler equations of an anharmonic chain, see
\cite{Sp13} for more details.
These are evolution equations of the conserved fields on a macroscopic scale. From \eqref{2.4} we infer that $r_j$ and $p_j$ are locally conserved.
As for any mechanical system also the local energy is conserved. The energy at site $j$ is
\begin{equation}\label{2.12}
e_j=\tfrac{1}{2}p^2_j + V(r_j)\,.
\end{equation}
Then
\begin{equation}\label{2.13}
\dot{e}_j= p_{j+1} V'(r_j)-p_j V'(r_{j-1})\,,
\end{equation}
hence the local energy current is $-p_j V'(r_{j-1})$. We collect the conserved fields as the $3$-vector
$\vec{g} = (g_1,g_2,g_3)$, 
\begin{equation}\label{2.14}
\vec{g}(j,t) = \big(r_j(t),p_j(t),e_j(t)\big) \,,
\end{equation}
$\vec{g}(j,0) = \vec{g}(j)$. Then
\begin{equation}\label{2.15}
\frac{d}{dt}\vec{g}(j,t) + \vec{\mathcal{J}}(j+1,t)  - \vec{\mathcal{J}}(j,t)=0 \,,
\end{equation}
where the local current functions are given by
\begin{equation}\label{2.16}
\vec{\mathcal{J}}(j) = \big( -p_j,-V'(r_{j-1}), - p_jV'(r_{j-1})\big)\,.
\end{equation}

Once the conserved fields are identified, the Euler equations follow from the assumption of local equilibrium.
More precisely, we introduce the microcanonical parameters $\ell$, $\mathsf{e}$ through
\begin{equation}\label{2.17}
\ell =\langle r_j\rangle_{p,\beta}\,,\quad \mathsf{e}=\langle e_j\rangle_{p,\beta}=\frac{1}{2\beta}+\langle V(r_j)\rangle_{p,\beta}\,.
\end{equation}
\eqref{2.17} defines $(p,\beta) \mapsto (\ell(p,\beta),\mathsf{e}(p,\beta))$, thereby the inverse map 
$(\ell, \mathsf{e}) \mapsto (p(\ell,\mathsf{e}),$ $  \beta(\ell,\mathsf{e}))$, and thus accomplishes the switch between
 the microcanonical variables $\ell, \mathsf{e}$ and the canonical variables $p, \beta$.
 Next let us choose an initial state, for which $p,\beta$, and mean velocity are slowly varying on the scale of the lattice.
 This induces a slow variation of stretch $\ell$, velocity $\mathsf{u}$, and total energy $\mathfrak{e} =\frac{1}{2} \mathsf{u}^2+ \mathsf{e}$.
Then  by averaging the fields in a local equilibrium state,  the microscopic conservation laws \eqref{2.15} turn into the Euler equations of an anharmonic chain as
\begin{equation}\label{2.18}
\partial_t\ell(x,t) +\partial_x \mathsf{j}_\ell(x,t) =0\,,\quad   \partial_t \mathsf{u}(x,t) +\partial_x \mathsf{j}_\mathsf{u}(x,t) =0\,,\quad   \partial_t \mathfrak{e}(x,t) +\partial_x \mathsf{j}_\mathfrak{e}(x,t) =0\,,
\end{equation}
where the hydrodynamic currents are given by
\begin{equation}\label{2.19}
\langle \vec{\mathcal{J}}(j)\rangle_{\ell,\mathsf{u},\mathfrak{e}} = \big(-\mathsf{u},p(\ell,\mathfrak{e}-\tfrac{1}{2}\mathsf{u}^2), \mathsf{u} p(\ell,\mathfrak{e}-\tfrac{1}{2}\mathsf{u}^2)\big)
= \vec{\mathsf{j}}
\end{equation}
with $p(\ell,\mathsf{e})$ defined implicitly through~\eqref{2.17}. By construction the slow variation refers to the particle label $j$. Hence ``$x$" in \eqref{2.18} stands for its continuum approximation.

Returning to the hard-point system with shoulder potential, one obtains
\begin{eqnarray}\label{2.20}
&&Z(p,\beta)  = \frac{1}{p\beta}\mathrm{e}^{-p\beta} \big(1 + \mathrm{e}^{-\beta}( \mathrm{e}^{p\beta /2} -1)\big)\,,\nonumber\\
&& \ell= -\frac{1}{\beta}\partial_p \log  Z(p,\beta)\,,\quad \mathsf{e} = \frac{1}{2\beta} + \frac{1}{Z(p,\beta)}
 \frac{1}{p\beta}\mathrm{e}^{-\beta-p\beta}( \mathrm{e}^{p\beta /2} -1)
 \,.\medskip
\end{eqnarray}

\paragraph{Biatomic chains.} We reintroduce the mass $m_j$ of the $j$-th particle and also a site-dependent
interaction potential $V_j$. Then the equations of motion for the chain become
\begin{equation}\label{2.20a}
\dot{r}_j = \frac{1}{m_{j+1}}p_{j+1} -  \frac{1}{m_j}p_j\,,\quad \dot{p}_j = V'_j(r_j)- V'_{j-1}(r_{j-1})\,.
\end{equation}
For a biatomic chain $m_j$ and $V_j$ have period 2 and hence the unit cell consists of two adjacent particles.
We normalize by $m_0$ and set $\kappa = m_1/m_0$. Then $m_j = 1$ for even $j$ and $m_j = \kappa$ for odd $j$. 
We consider the particular case, in which particles interact through the \emph{square-well potential}
\begin{equation}\label{2.21}
V_\mathrm{sw}(x) = 0  \mathrm{\hspace{4pt}for\hspace{4pt}} 0 < |x| < a\,,\quad
V_\mathrm{sw}(x) = \infty \mathrm{\hspace{4pt}otherwise\hspace{0pt}}\,.
\end{equation}
Then between collisions there is free motion with $\dot{p}_j = 0$. For $r_j = 0$ the incoming momenta are defined by 
$p_j - p_{j+1} > 0$ and for $r_j = a$ by $p_j - p_{j+1} < 0$. In either case
the collision rule reads
\begin{eqnarray}\label{2.22}
&&\hspace{0pt}p_j' = p_j + 2\,\frac{m_j p_{j+1} - m_{j+1}p_j}{m_j + m_{j+1}}\,,\nonumber\\
&&\hspace{0pt}p_{j+1}' = p_{j+1} - 2\, \frac{m_j p_{j+1} - m_{j+1}p_j}{m_j + m_{j+1}} \,.
\end{eqnarray}
Note that the transformation \eqref{2.22} depends only on the mass ratio $\kappa$.

Since there is zero potential energy,
\begin{equation}\label{2.24}
\mathsf{e} = \frac{1}{2\beta}
\end{equation}
and the pressure factorizes as
\begin{equation}\label{2.25}
a\beta p = h(\ell/a)\,,
\end{equation}
where $h$ is the inverse function to $y \mapsto y^{-1} - (\mathrm{e}^y -1)^{-1}$. Clearly,
length can be normalized such that $a=1$. This is then our model of hard-point particles
with alternating masses and square-well potential.

For the hard-point gas with merely alternating masses, we take the limit $a \to \infty$
which amounts to delete the option $r_j = a$ in the collision rules \eqref{2.22}. The pressure simplifies to
\begin{equation}\label{2.26}
\beta p = \frac{1}{\ell}\,.
\end{equation}

For the hydrodynamic equations one has to take into account that momentum and energy transfer in a collision
depend on the masses. Hence the currents \eqref{2.19} are modified to
\begin{equation}\label{2.27}
\big(-\frac{1}{\bar{m}}\mathsf{u},p,\frac{1}{\bar{m}}\mathsf{u}p\big)\,, \quad p = p(\ell,\mathfrak{e} - \frac{1}{2\bar{m}}\mathsf{u}^2)\,,
\end{equation}
where $\bar{m}$ stands for the average mass, $\bar{m} =(m_0 + m_1)/2$.


\section{Nonlinear fluctuating hydrodynamics}
\label{sec3}

A standing issue of statistical mechanics is to understand the long-time behavior of dynamical correlations for the chain in thermal equilibrium.
The modes with the longest life time will come from the locally conserved fields $\vec{g}(j,t) = \big(r_j(t),p_j(t),e_j(t)\big) $.
Hence one studies their correlations defined through
\begin{equation}\label{3.1}
S_{\alpha\alpha'}(j,t)=\langle g_{\alpha}(j,t) g_{\alpha'}(0,0)\rangle_{p,\beta} - \langle g_{\alpha}(j,t)\rangle_{p,\beta} \langle g_{\alpha'}(0)\rangle_{p,\beta}\,,
\end{equation}
$\alpha,\alpha'=1,2,3$. Note that by space-time stationarity  $\langle g_{\alpha}(j,t)\rangle_{p,\beta} = \langle g_{\alpha}(0)\rangle_{p,\beta}$ and
\begin{equation}\label{3.2}
S_{\alpha\alpha'}(j,t)= S_{\alpha'\alpha}(-j,-t)\,.
\end{equation}
Also at time $t=0$,
\begin{equation}\label{3.3}
S(j,0)=\delta_{j0} C
\end{equation}
which defines the static susceptibility matrix $C$. For the theory it is convenient to study directly the infinite one-dimensional lattice $\mathbb{Z}$, on which the correlations can spread forever. But MD is on a ring with $N$ sites and the dynamics is run only up to time $t_{\mathrm{max}}$, the first time when the two sound modes collide, i.e.,  $2c\, t_{\mathrm{max}}=N$ with $c$ the speed of sound.

In higher spatial dimensions the long-time properties of the correlation functions \eqref{3.1} are well captured by (linear) fluctuating hydrodynamics. This is a Gaussian fluctuation theory for the hydrodynamic fields. The drift part of the corresponding Langevin
equations is obtained by linearizing the Navier-Stokes equations around equilibrium and consists of the Euler flow term,
linear in $\partial_x$, and the dissipative transport terms, quadratic  in $\partial_x$. The noisy part is obtained by adding 
random currents with space-time white noise statistics to the systematic currents. These random currents model 
all the left out degrees of freedom from the exact conservation laws. The strength of the random currents is determined by the fluctuation dissipation theorem.  Fluctuating hydrodynamics
predicts diffusive broadening of the peaks. Such a behavior holds in dimension $d > 2$ and is well-known to break down in one dimension.
The minimal proposal in \cite{vB12,MS13} is to generalize to a nonlinear version, for which the Euler currents are kept up to second order in the deviation from equilibrium, while the dissipative part and the noise are taken from the linear theory. We give here a brief review with more details provided in \cite{Sp13}.
We fix the equilibrium parameters $p,\beta$ and denote the small deviations from equilibrium
by $\vec{u}(x,t) = (u_1(x,t),u_2(x,t),u_3(x,t))$. When
dissipation and noise are added, 
$\vec{u}(x,t)$ becomes a random field with zero average. By construction, the fluctuation field is governed by the Langevin equations
\begin{equation}\label{3.4}
\partial_t \vec{u}(x,t) +\partial_x \big(A \vec{u}(x,t) + \tfrac{1}{2} \langle
u(x,t),\vec{H}u(x,t)\rangle -\partial_x \tilde{D} \vec{u}(x,t) +\tilde{B}\vec{\xi}(x,t)\,\big)=0\,.
\end{equation}
Here the Euler currents have been expanded relative to the reference background $\vec{u}_0 = (\ell,0,\mathsf{e})$ 
up to second order as
\begin{equation}\label{3.5}
\mathsf{j}_\alpha (\vec{u}_0 + \vec{u}) = \mathsf{j}_\alpha (\vec{u}_0) + \sum_{\beta=1}^{3}\partial_{u_\beta} \mathsf{j}_\alpha (\vec{u}_0)u_\beta
 +\tfrac{1}{2} \sum_{\beta,\beta'=1}^{3}
\partial_{u_\beta} \partial_{u_{\beta'}} \mathsf{j}_\alpha (\vec{u}_0)u_\beta u_{\beta'}\,.
\end{equation}
This defines the $3\times 3$ linearization matrix $A$ and  the 
three-vector  of the Hessian matrices $\vec{H}$ of second derivatives,
\begin{equation}\label{3.6a}
A_{\alpha \beta}= \partial_{u_\beta} \mathsf{j}_\alpha\,, \quad  H^\alpha_{\beta\beta'} = 
\partial_{u_\beta} \partial_{u_{\beta'}} \mathsf{j}_\alpha\,.
\end{equation}
 $\vec{\xi}$ is Gaussian white noise with mean $0$ and covariance
\begin{equation}\label{3.6}
\langle\xi_\alpha(x,t) \xi_{\alpha'}(x',t')\rangle= \delta_{\alpha\alpha'} \delta(x-x') \delta(t-t')\,,
\end{equation}
where $\tilde{B}\tilde{B}\mathrm{^T}$ is the noise strength matrix with $^\mathrm{T}$ denoting transpose. 
The susceptibility matrix $C$ and the diffusion matrix $\tilde{D}$ satisfy the fluctuation-dissipation relation
\begin{equation}\label{3.7}
  \tilde{D}C+C\tilde{D}=\tilde{B}\tilde{B}^\mathrm{T}\,.
\end{equation}
If one had set $\vec{H} = 0$ in \eqref{3.4}, then this Langevin equation would agree with fluctuating hydrodynamics specialized to one dimension. In principle, one could include higher orders in the expansion. By power counting they are subdominant.
Of course, if quadratic coefficients vanish, one should study the effect of cubic terms.  Most likely, logarithmic corrections could result. But other features will be more important than such fine details.

We consider the stationary, mean zero solution  to \eqref{3.4}, again denoted by $\vec{u}(x,t)$. Then the claim is that for long times and large spatial scales
\begin{equation}\label{3.8}
\langle u_\alpha(x,t)u_\beta(0,0) \rangle \simeq S_{\alpha\beta}(j,t)
\end{equation}
with $x$ the continuum approximation for $j$.

For a single component Eq.~\eqref{3.4} is the stochastic Burgers equation, equivalently in its space integrated version, the one-dimensional Kadar-Parisi-Zhang equation \cite{KPZ86}. Multi-component KPZ type equations have been proposed before \cite{EK92,EK93}, however with degenerate, i.e., vanishing velocities. We refer to \cite{DBBR01} for pointing out the importance of 
distinct mode velocities. 

The linearization $A$ has the eigenvalues $-c,0,c$ corresponding to the left and right going sound peaks and the heat peak.
In Eq.~\eqref{3.4}, this linear term dominates all other terms. To better understand its role one has to make a linear transformation in component space, denoted by $R$,
such that $A$ becomes diagonal. In addition, as a convenient normalization, the transformed susceptibility matrix is required to be the unit matrix.
Both  conditions lead to 
\begin{equation}\label{3.9}
RAR^{-1}= \mathrm{diag}(-c,0,c)\,,\quad RCR\mathrm{^T}=1\,,
\end{equation}
which determine $R$ up to an overall sign. We set $\vec{\phi} = R\vec{u}$ and call $\vec{\phi} = (\phi_{-1},\phi_0,\phi_1)$ the normal modes. The transformed Langevin equations read
\begin{equation}\label{3.10}
\partial_t \phi_\alpha + \partial_x \big(c_\alpha \phi_\alpha + \langle\vec{\phi}, G^{\alpha}\vec{\phi}\rangle-\partial_x(D\phi)_\alpha+(B\xi)_\alpha\big)=0\,,
\end{equation}
$\alpha = -1,0,1$, where $ D = R \tilde{D}R^{-1}$ and $B = R\tilde{B}$ with noise strength $ B B^{\mathrm{T}} = 2 D$. The velocity of the $\alpha$-th normal mode is $c_\alpha$, $c_\sigma=\sigma c$, $c_0=0$, $\sigma=\pm1$.  The inner product $\langle\cdot,\cdot\rangle$ is in component space and the $G^\alpha$ matrix of coefficients stands for
\begin{equation}\label{3.11}
 G^\alpha = \tfrac{1}{2}\sum^3_{\alpha'=1}  R_{\alpha\alpha'}  (R^{-1})^{\mathrm{T}} H^{\alpha'}R^{-1}\,.
\end{equation}
As before, we have to consider the stationary process $\vec{\phi}(x,t)$ with mean zero, $\langle\vec{\phi}(x,t)\rangle = 0$,
satisfying  Eq.~\eqref{3.10}. The $\vec{\phi}$\hspace{1pt}-$\vec{\phi}$ correlations are defined by
\begin{equation}\label{3.12}
S^{\sharp\phi}_{\alpha\alpha'}(x,t) = \langle \phi_{\alpha}(x,t)  \phi_{\alpha'}(0,0) \rangle\,,
\end{equation}
where the superscript $^\sharp$ reminds of normal mode and $^\phi$ of the underlying stochastic process. 
The central claim is that, as $3\times 3$ matrices,
\begin{equation}\label{3.13}
RS(j,t)R^\mathrm{T} = S^\sharp(j,t) \simeq S^{\sharp\phi}(x,t) 
\end{equation}
on a mesoscopic scale. 

Eq.~\eqref{3.10} is a stochastic non-linear field theory and its two-point correlation cannot be readily computed.
We summarize the main findings up to now.\medskip\\
\emph{Diagonality}. By construction
 \begin{equation}\label{3.14}
S_{\alpha\alpha'}^{\sharp\phi}(x,0) = \delta_{\alpha\alpha'}\delta(x) \,.
\end{equation}
Using space-time stationarity and the conservation laws, one deduces the sum rule
\begin{equation}\label{3.14a}
\int dx\, S_{\alpha\alpha'}^{\sharp\phi}(x,t) = \int dx\, S_{\alpha\alpha'}^{\sharp\phi}(x,0) = \delta_{\alpha\alpha'} \,,
\end{equation}
but there is no reason for $S^{\sharp\phi}(x,t)$ to remain pointwise diagonal at later times. But the distinct velocities of the 
modes enforce such a behavior and,  in the one-loop mode coupling approximation to Eq.~\eqref{3.10},
the off-diagonal matrix elements are very small after some transient time. This is also seen in MD simulations and leads to 
\begin{equation}\label{3.14b}
S_{\alpha\alpha'}^{\sharp\phi}(x,t) \simeq  \delta_{\alpha\alpha'}f_\alpha(x,t)\,. 
\end{equation}
By  \eqref{3.14}, \eqref{3.14a}, the diagonal terms satisfy
\begin{equation}\label{3.15}
f_\alpha(x,0) = \delta(x)\,,\quad \int_{\mathbb{R}} dx\, f_\alpha(x,t) = 1\,. 
\end{equation}
For the physical fields
\begin{equation}\label{3.13a}
S_{\alpha\alpha} (j,t) \simeq \sum_{\alpha'=1}^{3}|(R^{-1})_{\alpha\alpha'}|^2f_{\alpha'}(x,t)\,. 
\end{equation}
Hence the Landau-Placzek ratios can be read off from the $R$ matrix. Generically, $S_{\alpha\alpha} (j,t)$
has three peaks located at $0$ (heat peak) and at $\pm ct$ (sound peaks). But for special parameter values some of the Landau-Plazcek ratios may vanish and less peaks are visible. \medskip\\
\emph{KPZ scaling, sound peaks}. Since the three modes have distinct propagation velocities, one expects Eq.~\eqref{3.10} to decouple into
three independent equations, each of which then has the structure of the noisy Burgers equation. Thus, if 
$G^\sigma_{\sigma\sigma} \neq 0$, one will have the KPZ scaling,
 \begin{equation}\label{3.14c}
f_\sigma(x,t)\cong (\lambda_\mathrm{s} t)^{-2/3} f_{\mathrm{KPZ}} \big((\lambda_\mathrm{s} t)^{-2/3}(x-\sigma ct)\big)\,.
\end{equation}
$f_{\mathrm{KPZ}}$ is the exact scaling function for the two-point correlation of the noisy Burgers equation,
see Appendix~\ref{secC}. According to KPZ scaling theory, the non-universal coefficient reads
\begin{equation}\label{3.15b}
\lambda_\mathrm{s} = |G^\sigma_{\sigma\sigma}| \, a_\mathrm{s}\,,\quad a_\mathrm{s} = 2\sqrt{2}\,,
\end{equation}
where we divided into the material parameter $G^\sigma_{\sigma\sigma}$ and the universal pure number 
$a_\mathrm{s}$. Of course, $a_\mathrm{s}$ depends on the convention for $f_{\mathrm{KPZ}}$.\medskip\\
\emph{L\'evy scaling, heat mode}. For anharmonic chains $G_{00}^0 = 0$, always. Thus the leading KPZ scaling 
\eqref{3.14c}
degenerates and one has to study the interaction between the modes. So far, this goal has been accomplished only on the level 
of mode-coupling, which leads to the prediction
\begin{equation}\label{3.16}
f_0(x,t) = (\lambda_\mathrm{h} t)^{-3/5}f_{\mathrm{L},5/3}((\lambda_\mathrm{h} t)^{-3/5} x)
\end{equation}
with $f_{\mathrm{L},\alpha}$ the symmetric $\alpha$-stable distribution,
also known as $\alpha$-L\'evy distribution,  see Appendix~\ref{secC}. As a result of previous numerical 
simulations \cite{CiDe05,DeDe07,ZaDe13},
and also confirmed here, $f_{\mathrm{L},5/3}$ seems to be the exact scaling function.
If so, one can use the scaling properties of non-linear fluctuating hydrodynamics to deduce that 
\begin{equation}\label{3.17}
\lambda_\mathrm{h} =   c^{-1/3}\lambda^{-2/3}_\mathrm{s} (G^0_{\sigma\sigma})^2 \, a_\mathrm{h}\,.
\end{equation}
As before, $a_\mathrm{h}$ is a pure number, not depending on the particular model. To determine  $a_\mathrm{h}$ one would have to rely on the exact solution
of some model in the same universality class. According to mode-coupling theory,
\begin{equation}\label{3.18}
a_\mathrm{h} = 4 \int_0^\infty \!\!ds \, s^{-2/3}\cos s \int_{\mathbb{R}}  dx f_{\mathrm{KPZ}}(x)^2 = 2 \sqrt{3}\, \Gamma\big(\tfrac{1}{3}\big) \int_{\mathbb{R}}  dx f_{\mathrm{KPZ}}(x)^2 \simeq 3.617\,.
\end{equation}
Physically one expects that there are no correlations propagating beyond the sound cone, which is confirmed 
in our simulations. Thus the Levy peak is cut off at the location of the sound modes.
\medskip\\
\emph{Even potential, zero pressure}. In principle, also $G^\sigma_{\sigma\sigma}$ could vanish implying that the prediction based on the noisy Burgers equation becomes invalid. One generic case for this to happen is $p=0$ and a potential
symmetric relative to some reference point. An example is $V_\mathrm{sw}$ with reference point $a/2$,
implying the non-KPZ value $\ell = a/2$.
Mode-coupling theory predicts the sound peaks to be diffusive,
\begin{equation}\label{3.19}
f_\sigma(x,t) = (\lambda_\mathrm{s} t)^{-1/2}f_{\mathrm{G}}((\lambda_\mathrm{s} t)^{-1/2}(x - \sigma ct))
\end{equation}
and the heat peak to be $\tfrac{3}{2}$-L\'evy,
\begin{equation}\label{3.20}
f_0(x,t) = (\lambda_\mathrm{h} t)^{-2/3}f_{\mathrm{L},3/2}((\lambda_\mathrm{h} t)^{-2/3}x)\,.
\end{equation}
Based on a recent exact solution for models in the same universality class \cite{BGJ14,JKO14},
and also confirmed by our simulations, the scalings \eqref{3.19}, \eqref{3.20} are expected to be the true asymptotic behavior.
From the self-similarity of non-linear fluctuating hydrodynamics one then deduces
\begin{equation}\label{3.21}
\lambda_\mathrm{h} =   c^{-1/2}\lambda^{-1/2}_\mathrm{s} (G^0_{11})^2 a_\mathrm{h}\,.
\end{equation}
The exact solution implies
\begin{equation}\label{3.22}
 a_\mathrm{h} =  4 \int_0^\infty \!\!ds s^{-1/2} \cos s  \int_{\mathbb{R}}  dx f_{\mathrm{G}}(x)^2 = \sqrt{2}\,,
 \end{equation}
 which happens to agree with the mode-coupling computation.\medskip\\
\textit{Remark}. Eq.~(4.11) of \cite{Sp13} should read $\exp[-|2\pi k|^{5/3}\lambda_\mathrm{h}t ]$ and consequently $\lambda_\mathrm{h}$ of
Eq.~(4.12) has to be multiplied by $(2\pi)^{-5/3}$. On the same footing, in Eq.~(4.18) it should read $\exp[-|2\pi k|^{3/2}\lambda_\mathrm{h}t ] $
and consequently $\lambda_\mathrm{h}$ of
Eq. (4.19) has to be multiplied by $(2\pi)^{-3/2}$.\medskip

For the hard-point systems under study, the free energy and the Euler currents have been provided already and this allows for the computation of the non-universal constants, at least in principle. However, for equal masses with square shoulder potential,
while \eqref{2.20} looks still simple, to compute, say, $G$ as a function of $p,\beta$ turns out to be cumbersome. 
Therefore we rely on a \textsf{Mathematica} code, which computes all coefficients numerically. 
For alternating masses with square-well potential, since the pressure factorizes, all coefficients are
expressed in terms of $h$ and its derivatives. To have at least one explicit example, we provide the details
in Appendix~\ref{secA}.

Note that nonlinear fluctuating hydrodynamics makes predictions 
in essence independent of the specific value of the mass ratio $\kappa$. So we could set $\kappa = 1$. But for the mechanical system
this amounts to a mere relabeling. Thus  in our derivations implicitly we have assumed that the dynamics is sufficiently chaotic and that the system has no other conservation laws than the three  listed already.

\section{Molecular dynamics simulations}
\label{sec4}
Nonlinear fluctuating hydrodynamics is based on several assumptions. To find out about the accuracy of the theory  
one has to rely on MD simulations, which have been carried out for all three models, in each case for a single choice of parameters.
The lattice size is always $N=4096$. For given initial conditions the dynamics is obtained by iterating collision 
after collision. As an example, for the shoulder potential at our choice of parameters there are approximately $1200$ collisions per particle up to the maximal time 
$t_{\max} = 1024$. In our implementation we use an ``event table'' consisting of 8192 ($= 2^{13}$) time slots, each of which covers the interval $1/8192$. At the beginning of the simulation, pairwise collision events are determined from the positions and momenta of neighboring particles. Each anticipated collision event is stored in the time slot covering the event timing modulo $1$. A time slot can store more than one event, but the time interval of a slot is chosen such that there is typically only one event per slot. Conceptually, the event table resembles a hash table, with the event time serving as index. During the actual simulation, the time slots are cyclically traversed one after another: we pick the (closest in time) event from the current time slot and update the positions and momenta of the particle pair associated with the event to the time point immediately after the collision. Since momenta have changed, the predictions for the neighboring particles have to be revised, and associated collision events are moved to the time slot covering the newly predicted collision time. The phase space functions defining $S(j,t)$ are averaged over all lattice sites and recorded for all $j$ and at times $t = 256$, $512$, $1024$. We use fast Fourier transformation to accelerate this step. A simulation for the shoulder potential at our parameters takes approximately $1.5\,\mathrm{s}$ on a commodity laptop computer. The scheme is repeated $10^7$ times with initial conditions sampled by means of a random number generator from the i.i.d.~distribution defined at and above Eq.~\eqref{2.10}.
In the last step we perform the linear transformation $RS(j,t)R^\mathrm{T} = S^{\sharp}(j,t)$. In fact, as striking qualitative prediction,
this matrix should be diagonal in good approximation. Indeed, the off-diagonal matrix elements have size less than $4\%$ of the diagonal entries, and in the figures below we only show the diagonal entries $S^{\sharp}_{\alpha\alpha}(j,t)$, $\alpha = \pm 1,0$.
By symmetry $S^{\sharp}_{\alpha\alpha}(j,t)= S^{\sharp}_{-\alpha-\alpha}(-j,t)$ and only one sound peak
needs to be plotted.

Having obtained the numerical peak, $f^\mathrm{num}_\alpha$, one has to compare with the theoretical prediction
$f^\mathrm{th}_\alpha$. Since by construction the area under each peak equals $1$ and since the peaks turn out to be positive, it is natural to use the $L^1$-norm as a numerical value for the distance between $f^\mathrm{num}_\alpha$
and $f^\mathrm{th}_\alpha$. As \textit{only free parameter} we adopt the linear scale
 and minimize the expression 
\begin{equation}
\label{4.1}
\sum_{j=1}^{N} \big| f^\mathrm{num}_\alpha(j,t) -
(\lambda t)^{-\gamma_{\alpha}} f^\mathrm{th}_\alpha((\lambda t)^{-\gamma_{\alpha}}(j -c_\alpha t))\big| 
\end{equation}
with respect to $\lambda > 0$ for fixed $t$. Here $c_\alpha$ is the velocity of mode $\alpha$
and $\gamma_{\alpha}$ is the theoretical scaling exponent. We record the minimal $L^1$-distance
and the respective value of $\lambda$. Obviously, there is some level of arbitrariness in our choice.

We discuss the MD results for each model separately. The transformation matrix $R$ and the nonlinear couplings $G$ are listed in Appendix \ref{secB}. Our conventions for the scaling functions can be found in Appendix \ref{secC}.

\begin{figure}[!t]
\centering
\subfloat[overview]{
\includegraphics[width=0.33\columnwidth]{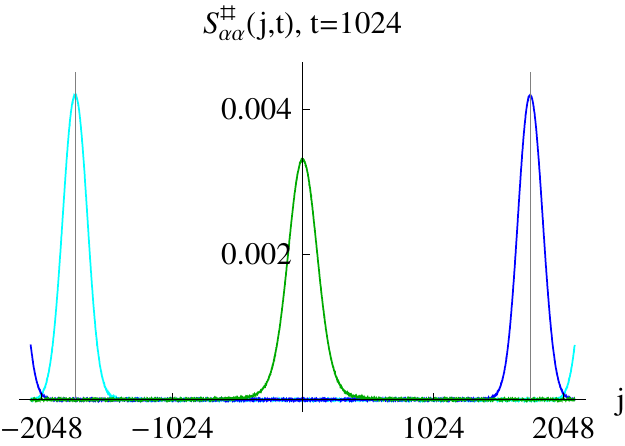}}
\subfloat[heat, $\lambda = 1.624$]{
\includegraphics[width=0.33\columnwidth]{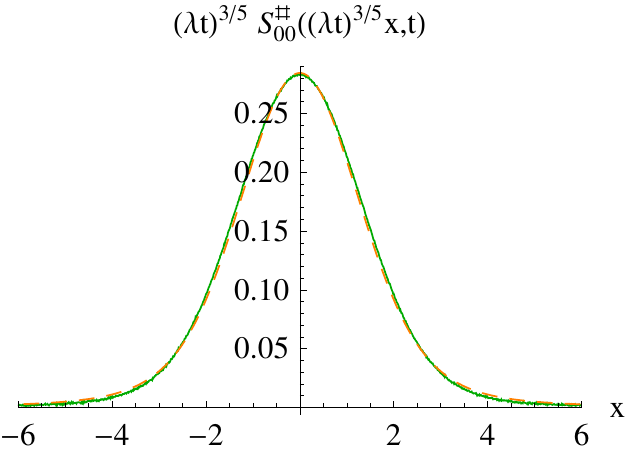}}
\subfloat[sound, $\lambda = 1.442$]{
\includegraphics[width=0.33\columnwidth]{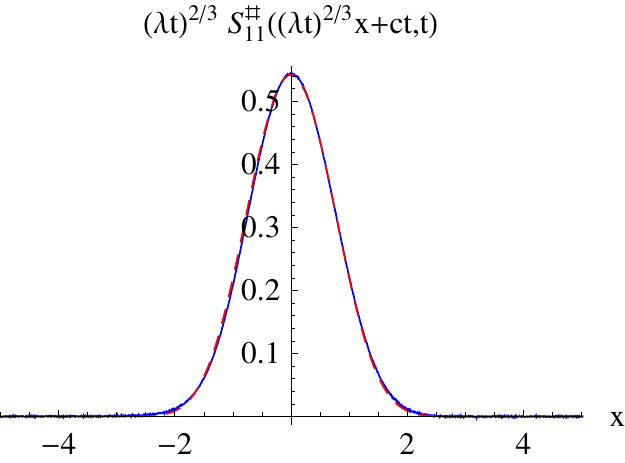}}
\caption{(Color online) MD simulation of an equal mass chain with shoulder potential as defined in Eq.~\eqref{2.2} and parameters $N = 4096$, $p = 1.2$, $\beta = 2$, at $t = 1024$. (a) Diagonal matrix entries, $S^{\sharp}_{\alpha\alpha}(j,t)$, of the two-point correlations. The gray vertical lines show the sound speed predicted from theory. The tails of the sound peaks reappear on the opposite side due to periodic boundary conditions. (b) Rescaled heat and (c) right sound peak. The theoretical scaling exponents are used and $\lambda$ is fitted numerically to minimize the $L^1$-distance between simulation and prediction. The dashed orange curve is the predicted $\tfrac{5}{3}$-L\'evy distribution $f_{\mathrm{L},5/3}$ and the dashed red curve shows $f_{\mathrm{KPZ}}$.}
\label{fig:shoulder_overview}
\end{figure}

\paragraph{Shoulder potential.}
This is an equal mass chain with potential \eqref{2.2}.
The parameters are $p = 1.2$ and $\beta = 2$, yielding the sound speed $c = 1.743$
and the average stretch $\langle r_j \rangle = 1.246$. In Fig.~\ref{fig:shoulder_overview}a the three peaks are superimposed. For the correlations of the physical fields each peak comes with a weight,
see Eq.~\eqref{3.13a}. For the stretch correlations
the weights are $0.082 : 0.065 : 0.082$, for the momentum correlations $0.25 : 0 : 0.25$, and for the energy correlations
$0.119 : 0.07 : 0.119$. In Fig.~\ref{fig:shoulder_overview}b,c the scaled heat and sound peaks are compared with the theoretical predictions. We note that the deviation from the theoretical shape is fairly small, but the non-universal $\lambda$-coefficients are still dropping in time. The prediction for $a_\mathrm{s}$ is based on decoupling, which is expected  to be exact.
The theoretical value is $a_\mathrm{s} = 2\sqrt{2} \simeq 2.828$, to be compared with the $t = 1024$ molecular dynamics value of 3.936, indicating that the simulation has not yet reached
the asymptotic regime. The theoretical $\tfrac{5}{3}$-L\'evy distribution of the heat peak is based on mode-coupling. 
From this perspective, it is not even sure that the true scaling function is given by $\tfrac{5}{3}$-L\'evy. But our simulations, and also earlier results
\cite{CiDe05,ZaDe13}, support a symmetric stable distribution with exponent $\tfrac{5}{3}$.

We record the still drifting non-universal coefficients in Table~\ref{tab:shoulder_lambda}, together with  the $L^1$ distance defined in Eq.~\eqref{4.1}.  For ease of comparison we provide 
the universal coefficients $a_\mathrm{s}$,  $a_\mathrm{h}$. The theory value for $a_\mathrm{s}$ is exact, whereas  $a_\mathrm{h}$ employs mode-coupling theory. In Table~\ref{tab:shoulder_lambda} the theory value for 
$\lambda_\mathrm{h}$ is based on the exact value of $\lambda_\mathrm{s}$. One could argue that instead  the measured value of $\lambda_\mathrm{s}$ at the same time should be used. This will make the comparison slightly less favorable.
It is remarkable that the empirical and theoretical values are in reasonable agreement, despite 
the system not yet having reached the asymptotic regime. To have a quantitative test, one would have to simulate for longer times and, consequently, with larger lattices. 
\begin{table}[!ht]
\centering
\begin{tabular}{l|cccc|c|ccc}
shoulder potential & $t = 256$ & $t = 512$ & $t = 1024$ & theory & & $t = 1024$ & theory\\
\hline
heat: \hspace{20pt}$\lambda_{\mathrm{h}}$ & 1.807 & 1.713 & 1.624 & 1.711 mc & $a_\mathrm{h}$ & 3.433 & 3.617 mc\\
\hspace{30pt}$L^1$-distance & 0.051 & 0.047 & 0.042 & & & & \\
\hline
sound: \hspace{20pt}$\lambda_{\mathrm{s}}$ & 1.735 & 1.575 & 1.442 & 1.036 & $a_\mathrm{s}$ & 3.936 & 2.828 \\
\hspace{30pt}$L^1$-distance & 0.043& 0.037 & 0.032 & & & & \\
\hline
\end{tabular}
\caption{Numerically fitted non-universal coefficients from the shoulder potential simulation of Fig.~\ref{fig:shoulder_overview}, and the corresponding $L^1$-distance to the theoretically predicted stable distribution $f_{\mathrm{L},5/3}$ for the heat peak and $f_{\mathrm{KPZ}}$ for the sound peak.}
\label{tab:shoulder_lambda}
\end{table}

\begin{figure}[!t]
\centering
\subfloat[$t = 256$]{
\includegraphics[width=0.33\columnwidth]{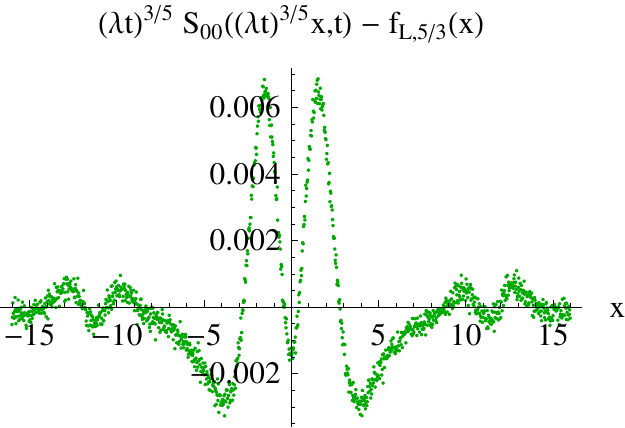}}
\subfloat[$t = 512$]{
\includegraphics[width=0.33\columnwidth]{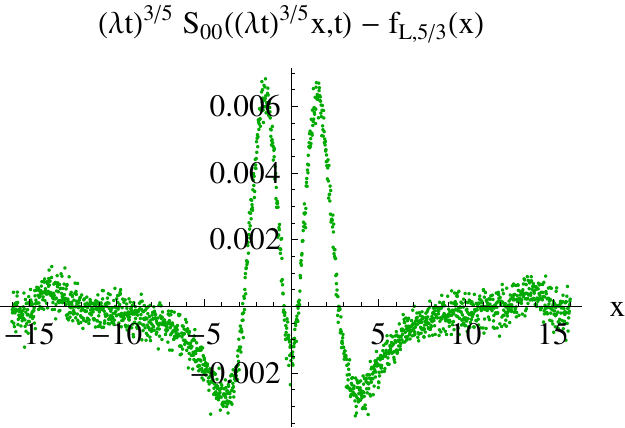}}
\subfloat[$t = 1024$]{
\includegraphics[width=0.33\columnwidth]{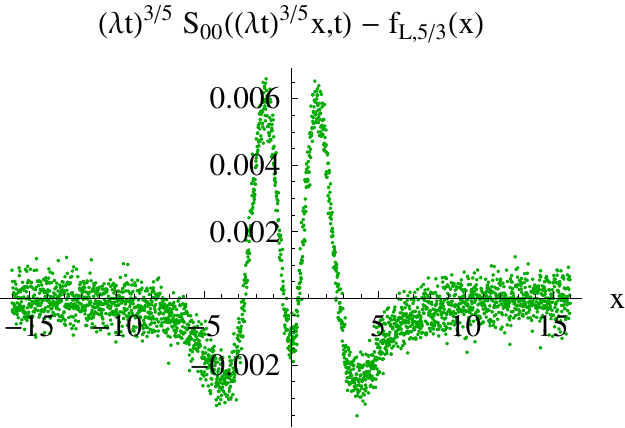}} \\
\subfloat[$t = 256$]{
\includegraphics[width=0.33\columnwidth]{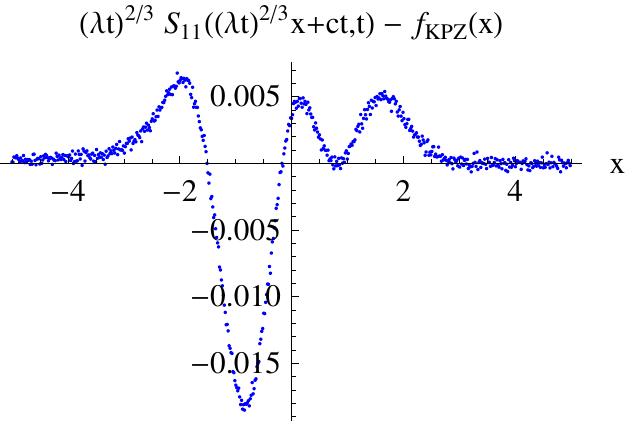}}
\subfloat[$t = 512$]{
\includegraphics[width=0.33\columnwidth]{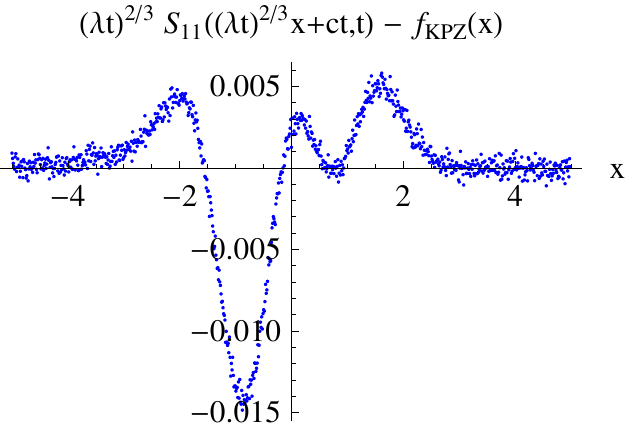}}
\subfloat[$t = 1024$]{
\includegraphics[width=0.33\columnwidth]{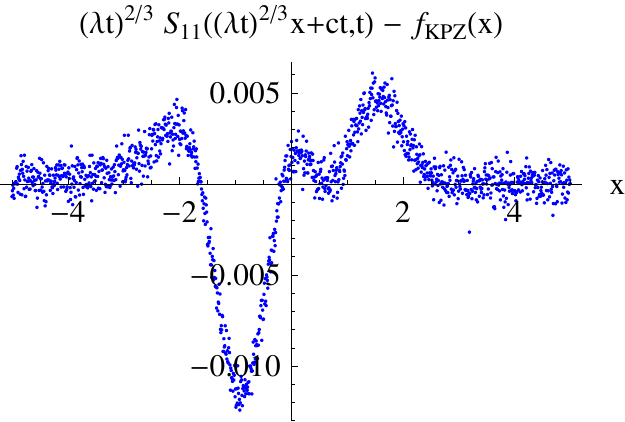}}
\caption{(Color online). Difference between, respectively,  the heat and right sound peaks obtained from the MD simulation with shoulder potential and the theoretical prediction at optimal $\lambda$, as listed in Table~\ref{tab:shoulder_lambda}. The notches around $\lvert x \rvert \simeq 12$ in (a) are due to feedback from the sound modes.}
\label{fig:shoulder_diff}
\end{figure}

\begin{figure}[!t]
\centering
\subfloat[shoulder, heat]{
\includegraphics[width=0.33\columnwidth]{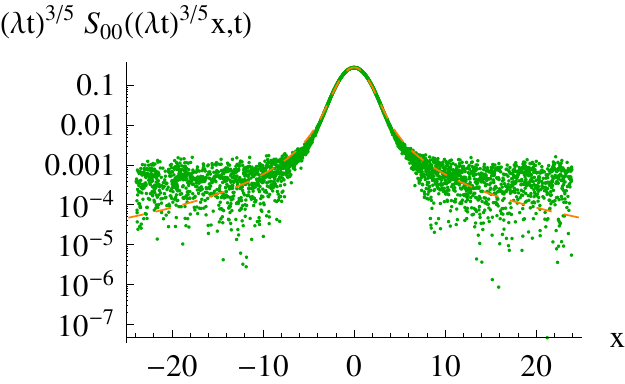}}
\subfloat[hard-point gas, heat]{
\includegraphics[width=0.33\columnwidth]{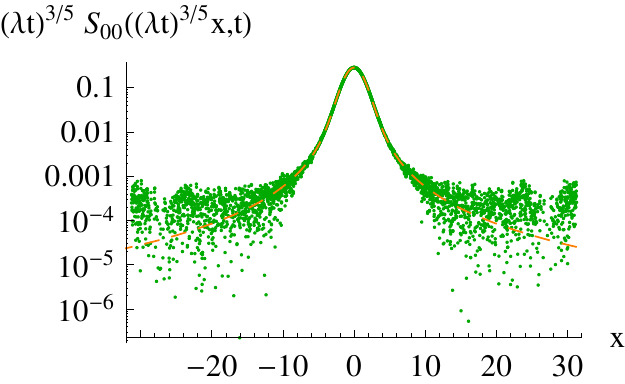}}
\subfloat[square-well, heat]{
\includegraphics[width=0.33\columnwidth]{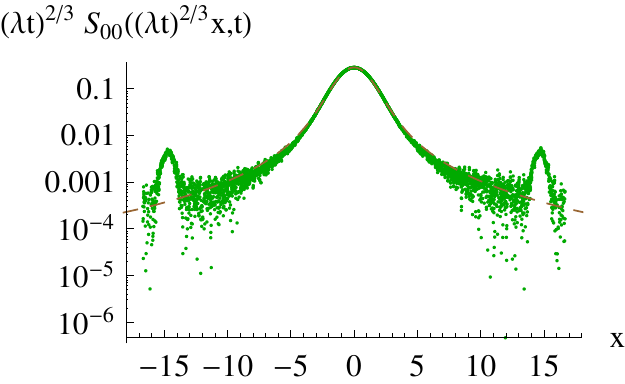}} \\
\subfloat[shoulder, sound]{
\includegraphics[width=0.33\columnwidth]{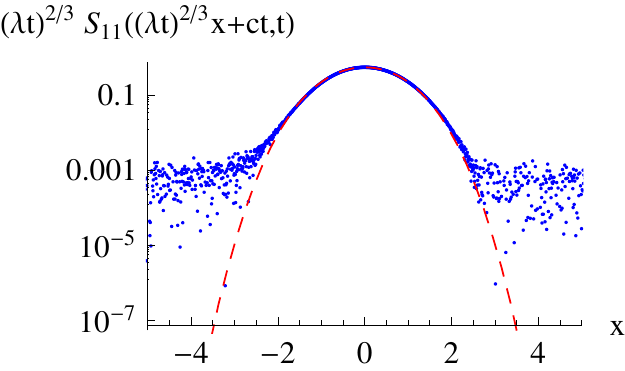}}
\subfloat[hard-point gas, sound]{
\includegraphics[width=0.33\columnwidth]{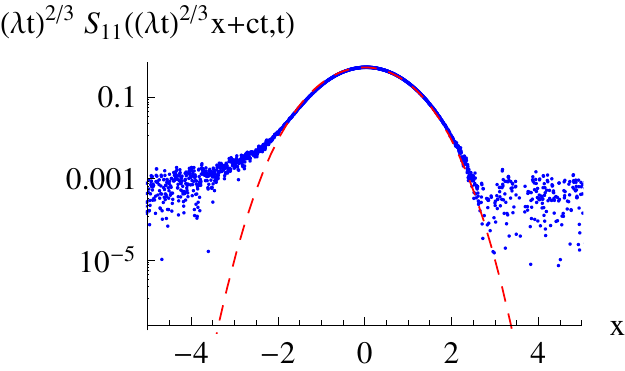}}
\subfloat[square-well, sound]{
\includegraphics[width=0.33\columnwidth]{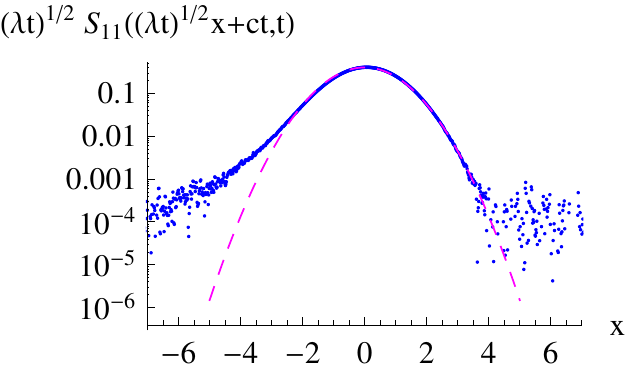}}
\caption{(Color online). Logarithmic plot of the heat and right sound peaks for all three models, at $t = 1024$. The dashed orange curve in (a) and (b) is the $f_{\mathrm{L},5/3}$ and the dashed brown curve in (c) the $f_{\mathrm{L},3/2}$. The dashed red curve in (d) and (e) shows $f_{\mathrm{KPZ}}$, and the magenta dashed curve in (f) is 
the Gaussian $f_{\mathrm{G}}$.}
\label{fig:log}
\end{figure}

Given the good fit in Fig.~\ref{fig:shoulder_overview}, more details are provided by  plotting the difference between the simulation data and the theoretical fit at optimal $\lambda$, see Fig.~\ref{fig:shoulder_diff} with a logarithmic plot 
provided in Fig.~\ref{fig:log}. Even for this difference, the change from the earliest time, $t =256$, to the latest one, $t = 1024$, is not particularly pronounced.

We also simulated the dynamics with an attractive potential, for which \eqref{2.2} is modified such that $V_\mathrm{sh}^{-}(x) = -1$ for $\frac{1}{2} < |x| < 1$. The parameters are fixed as $\beta = \frac{2}{5}$ and $p = \frac{3}{2}$, with a corresponding sound velocity $c = 1.745$. The coupling matrix $G^{0}$ hardly changes, while $G^{1}$ is roughly doubled. This leads to broader sound peaks and thus a stronger interaction between the peaks. The heat peak has the same error bars as in case of the repulsive potential.  At the longest time the sound peaks still have a slight asymmetry, increasing the $L^1$ distance by a factor of $3$. The attractive potential, at the given parameters,
seems to have a considerably slower convergence. Indicative are $a_{\mathrm{h}} = 32.447$, $a_{\mathrm{s}} = 12.413$, both at $t = 1024$ and corresponding to $\lambda_{\mathrm{h}} = 7.209$, $\lambda_{\mathrm{s}} = 9.449$, which deviate even further from the theoretical values.

\begin{figure}[!t]
\centering
\subfloat[$t = 256$]{
\includegraphics[width=0.33\columnwidth]{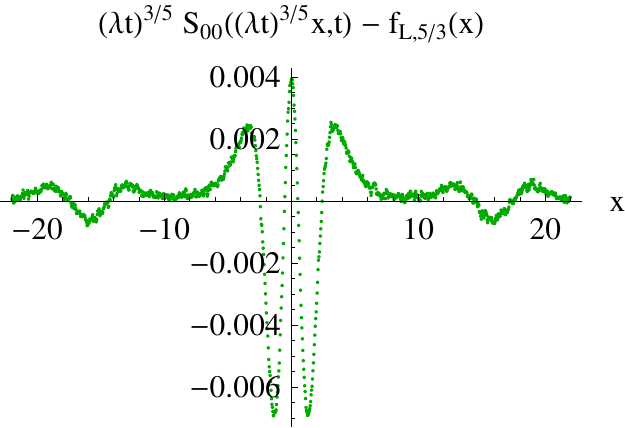}}
\subfloat[$t = 512$]{
\includegraphics[width=0.33\columnwidth]{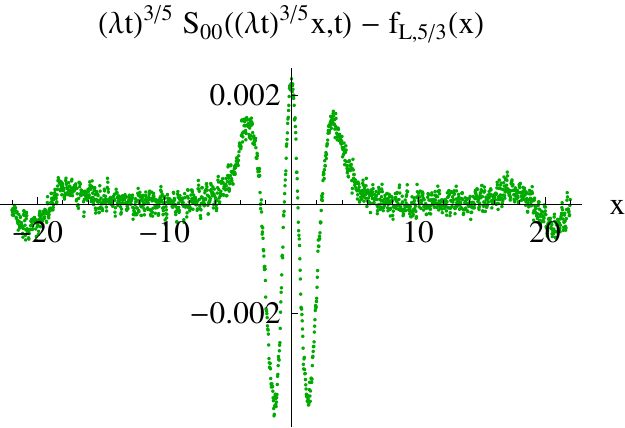}}
\subfloat[$t = 1024$]{
\includegraphics[width=0.33\columnwidth]{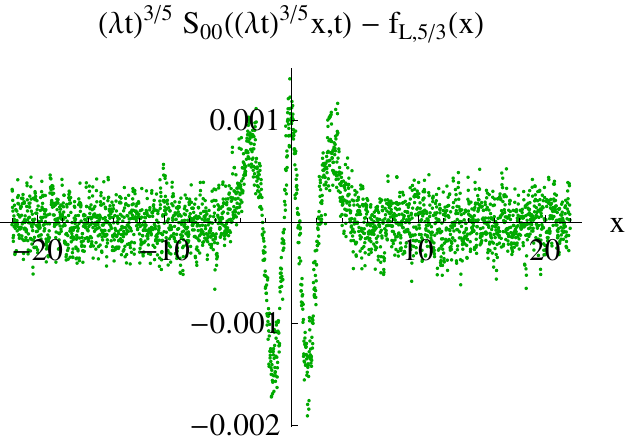}} \\
\subfloat[$t = 256$]{
\includegraphics[width=0.33\columnwidth]{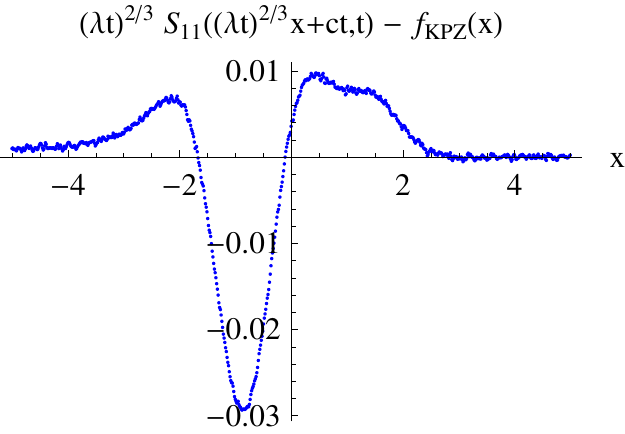}}
\subfloat[$t = 512$]{
\includegraphics[width=0.33\columnwidth]{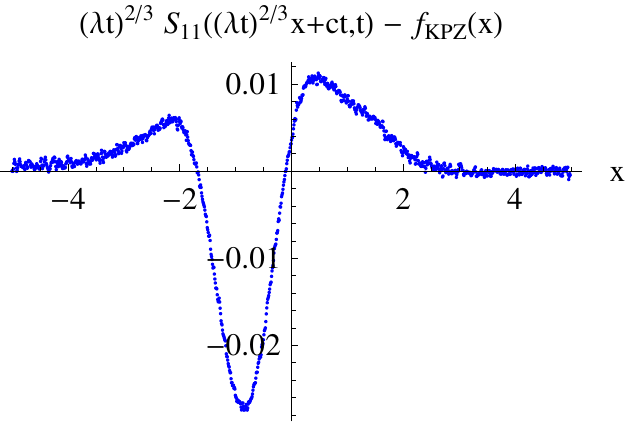}}
\subfloat[$t = 1024$]{
\includegraphics[width=0.33\columnwidth]{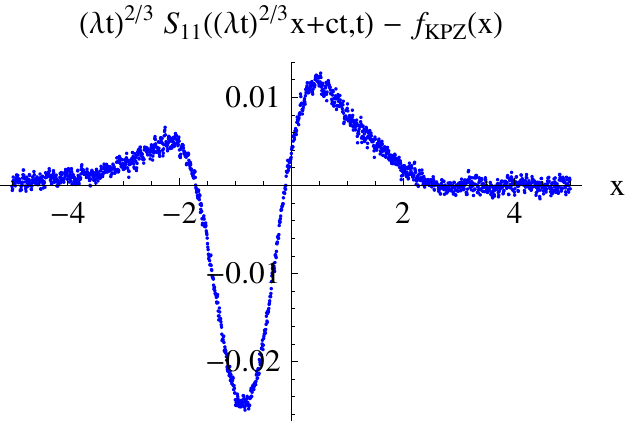}}
\caption{(Color online). Difference between, respectively,  the heat and right sound peaks obtained from the MD simulation with alternating masses $m_0 = 1$, $m_1 = 3$, and the theoretical prediction at optimal $\lambda$. Each $\lambda$ is fitted numerically to minimize the $L^1$-distance, see Table~\ref{tab:altern_mass_lambda}. Note the notches at $\lvert x \rvert \simeq 16$ in (a) resulting from feedback of the sound modes.}
\label{fig:altern_mass_diff}
\end{figure}

\paragraph{Hard-point gas with alternating masses.}
For biatomic chains, the unit cell consists of two adjacent particles. To allow for direct comparison with monoatomic chains, we average the two-point correlations according to
\begin{equation}
\label{eq:Sunit2}
\tilde{S}_{\alpha\alpha'}(j,t) = \tfrac{1}{4} \big( 2\, S_{\alpha\alpha'}(j,t) + S_{\alpha\alpha'}(j-1,t) + S_{\alpha\alpha'}(j+1,t) \big).
\end{equation}
Omitting such average, the two-point correlations would have a pronounced period of $2$. The parameters of the hard-point gas are alternating masses $m_0 = 1$, $m_1 = 3$ and $p = 2$, $\beta = 1/2$,
yielding a sound speed of $c_{\bar{m}} = \sqrt{3}$.
The peak structure is comparable to Fig.~\ref{fig:shoulder_overview}. For the stretch correlations
the weights are $\frac{1}{6}:\frac{2}{3}:\frac{1}{6}$, for the momentum correlations $2:0:2$, and for the energy correlations
$\frac{2}{3}:\frac{2}{3}:\frac{2}{3}$. The difference between the rescaled sound and heat peaks and the theoretical prediction is displayed in Fig.~\ref{fig:altern_mass_diff}.
We record the still drifting non-universal coefficients in Table~\ref{tab:altern_mass_lambda} together with the
prediction for the universal coefficients. Note that, despite different material parameters, the accuracy is comparable to the
chain with shoulder potential. 
\begin{table}[!ht]
\centering
\begin{tabular}{l|cccc|c|ccc}
hard-point gas  & $t = 256$ & $t = 512$ & $t = 1024$ & theory & & $t = 1024$ & theory\\
\hline
heat:\hspace{20pt}$\lambda_{\mathrm{h}}$ & 0.982 & 1.021 & 1.039 & 0.949 mc & $a_\mathrm{h}$ & 3.961 & 3.617 mc \\
\hspace{30pt}$L^1$-distance & 0.046 & 0.027 & 0.015 & & & & \\
\hline
sound:\hspace{20pt}$\lambda_{\mathrm{s}}$ & 2.540 & 2.482 & 2.421 & 2 & $a_\mathrm{s}$ & 3.424 & 2.828 \\
\hspace{30pt}$L^1$-distance& 0.063 & 0.057 & 0.053 & & & \\
\hline
\end{tabular}
\caption{Numerically fitted non-universal coefficients for the hard-point gas with alternating masses $m_0 = 1$, $m_1 = 3$, and the corresponding $L^1$-distance to the  $\tfrac{5}{3}$-L\'evy distribution 
$f_{\mathrm{L},5/3}$ for the heat peak and KPZ scaling function $f_{\mathrm{KPZ}}$ for the sound peak.}
\label{tab:altern_mass_lambda}
\end{table}

\pagebreak

\paragraph{Square-well potential with alternating masses at zero pressure.}
The universality classes of nonlinear fluctuating hydrodynamics depend on the vanishing of some of the leading couplings $G^\alpha_{\alpha'\alpha'}$. For anharmonic chains, $G^0_{00} =0$ always. The make some other leading coefficient  vanish is not so easily achieved, except for a symmetric potential at zero pressure. A specific example is 
 the square-well potential  at zero pressure. While the overall appearance looks similar, one can test a universality class different from the previous two examples.
The square-well potential is defined in Eq.~\eqref{2.21}. We use the maximal distance $a = 1$,
alternating masses $m_0 = 1$, $m_1 = 3$, and $p = 0$, $\beta = 2$, yielding a sound speed of $c_{\bar{m}} = \sqrt{3}$. The peak structure is comparable to Fig.~\ref{fig:shoulder_overview}. For the stretch correlations
the weights are $\frac{1}{24}:0:\frac{1}{24}$, for the momentum correlations $\frac{1}{2}:0:\frac{1}{2}$, and for the energy correlations
$0:\frac{1}{8}:0$. 
We record the still drifting non-universal coefficients in Table~\ref{tab:max_dist_lambda}. For the theoretical prediction,
as input for \eqref{3.21} we use the measured value of $\lambda_{\mathrm{s}}$ at $t = 1024$. $\lambda_{\mathrm{s}}$
is a regular transport coefficient not covered by our version of fluctuating hydrodynamics.
\begin{table}[!ht]
\centering
\begin{tabular}{l|ccc|c|cc}
square-well, $a = 1$ & $t = 256$ & $t = 512$ & $t = 1024$ & & $t = 1024$ & theory\\
\hline
heat:\hspace{20pt}$\lambda_{\mathrm{h}}$ & 1.502 & 1.410 & 1.324 & $a_\mathrm{h}$ & 2.423 & 1.414  \\
\hspace{30pt}$L^1$ distance & 0.054 & 0.048 & 0.042 & & & \\
\hline
sound:\hspace{20pt}$\lambda_{\mathrm{s}}$ & 3.902 & 4.146 & 4.348 & & & \\
\hspace{30pt}$L^1$-distance & 0.058 & 0.056 & 0.051 & & \\
\hline
\end{tabular}
\caption{Numerically fitted non-universal coefficients for the simulation with alternating masses $m_0 = 1$, $m_1 = 3$ and square-well potential, and the corresponding $L^1$-distance to the $\tfrac{3}{2}$-L\'evy distribution 
$f_{\mathrm{L},3/2}$ for the heat peak and Gaussian $f_{\mathrm{G}}$ for the sound peak. The theory value uses the current numerical value of $\lambda_{\mathrm{s}}$.}
\label{tab:max_dist_lambda}
\end{table}\\

\begin{figure}[!t]
\centering
\subfloat[$t = 256$]{
\includegraphics[width=0.33\columnwidth]{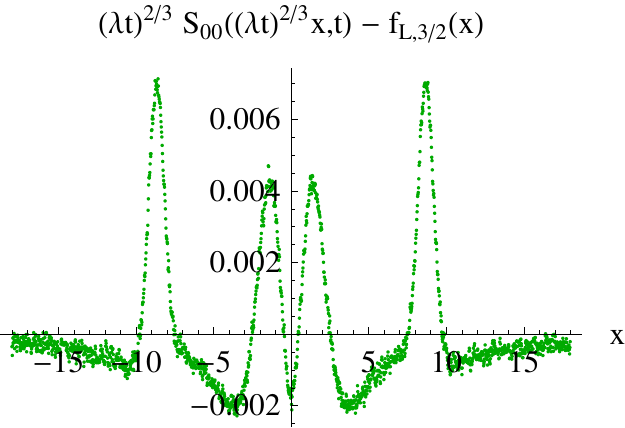}}
\subfloat[$t = 512$]{
\includegraphics[width=0.33\columnwidth]{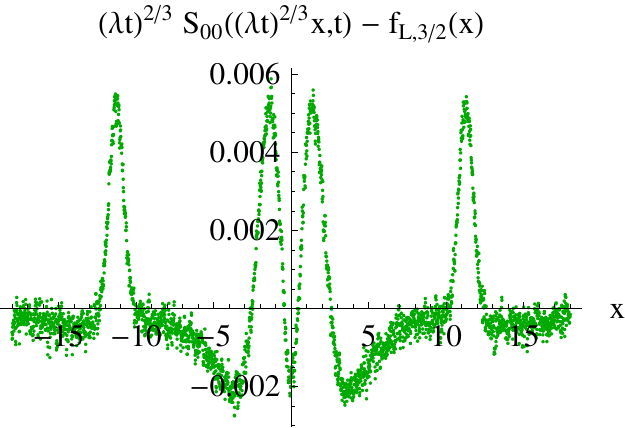}}
\subfloat[$t = 1024$]{
\includegraphics[width=0.33\columnwidth]{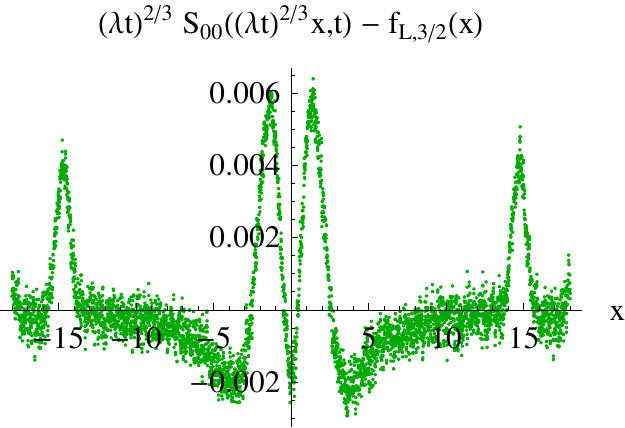}} \\
\subfloat[$t = 256$]{
\includegraphics[width=0.33\columnwidth]{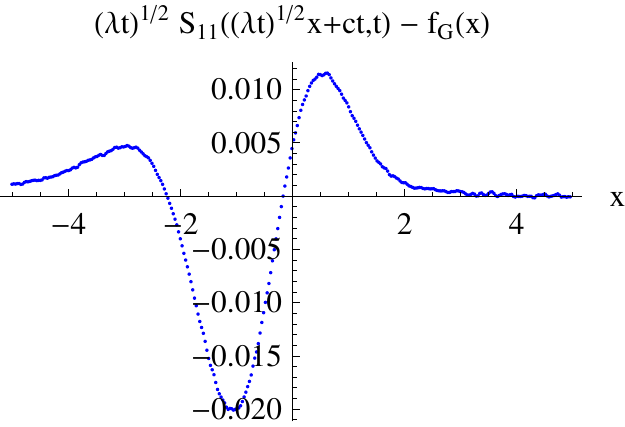}}
\subfloat[$t = 512$]{
\includegraphics[width=0.33\columnwidth]{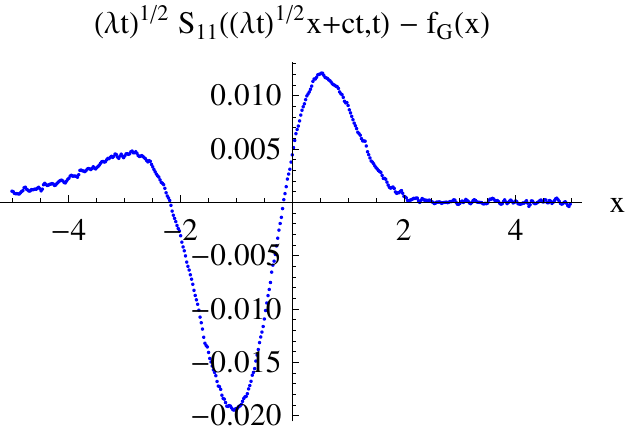}}
\subfloat[$t = 1024$]{
\includegraphics[width=0.33\columnwidth]{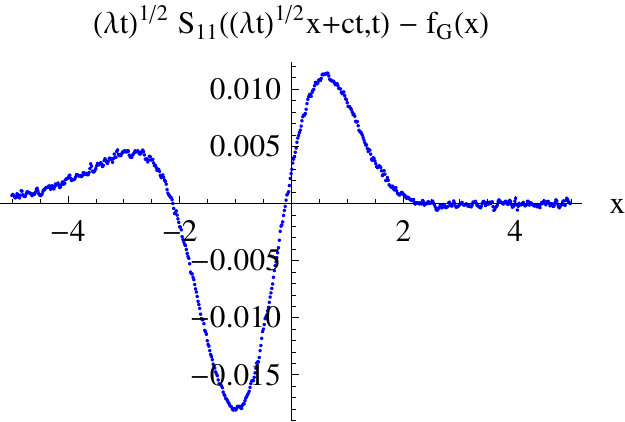}}
\caption{(Color online). Difference between the, respectively, rescaled heat and right sound peaks obtained
from the MD simulation with maximal distance $a = 1$ and alternating masses, and the theoretical prediction
at optimal $\lambda$. Note that the exponents and asymptotic functions are different from the previous cases. The fitted $\lambda$ values are provided in Table~\ref{tab:max_dist_lambda}. In the top row, the feedback from the sound modes to the heat mode is clearly visible.}
\label{fig:max_dist_rescale}
\end{figure}

The difference between the rescaled sound and heat peaks and the theoretical prediction is displayed in Fig.~\ref{fig:max_dist_rescale}. It is also instructive to have a logarithmic plot of the simulation data.
In Fig.~\ref{fig:log},  for each of the three models, we only show the longest time. The fit is for optimal $\lambda$. The asymmetry of the sound peak is still visible with a slightly slower decay towards the heat peak.

\section{Conclusions}
\label{sec5}

For a few 
anharmonic chains with hard collisions, our MD simulations support the predictions from nonlinear fluctuating hydrodynamics. Two of the models have sound peaks satisfying KPZ scaling and a heat peak which scales
as the symmetric $\tfrac{5}{3}$-L\'evy distribution. The square-well potential chain  is anomalous 
at the fine-tuned parameters $p=0$, $\ell =a/2$, in having  two diffusive sound peaks and a symmetric $\tfrac{3}{2}$-L\'evy heat peak. Of course, it would be of interest to expand the evidence by investigating
FPU chains and possibly one-dimensional classical fluids.

The, to us, most surprising discovery  is the precision at which the peaks fit the predicted scaling functions.
In particular, we add to the evidence that the stable law with exponent $\tfrac{5}{3}$ is indeed the exact scaling function.
 The peaks attain their theoretical shape already for fairly short times.
 On the other side, the non-universal coefficients $\lambda_{\mathrm{s}}$ and 
$\lambda_{\mathrm{h}}$ are still slowly drifting on the appropriate self-similar scale. On the 
sizes and times accessible by the simulation, the $\lambda$ coefficients have not reached a limiting  value.
The deviation of  
$\lambda_{\mathrm{s}}$, $\lambda_{\mathrm{h}}$ from their theoretical value is significant, but one could believe that eventually the predicted asymptotics  will be reached.  
Our observation, if true in more generality, would shed some light on earlier discrepancies in determining scaling exponents. It is like averaging over systems with the same scaling exponent but distinct non-universal parameters.
In the same spirit we point out that in Figs. \ref{fig:shoulder_diff} to \ref{fig:log} the sound peaks 
show a slight asymmetry and the heat peak has bumps resulting from the interaction with the sound mode.
We conjecture that these are transient effects which will disappear for longer times and correspondingly larger 
system size.

In the literature there are contributions which point to similar conclusions. We mention the early measurement of the 
total energy-energy current correlations  with a decay as $t^{-2/3}$ \cite{GrNa02}. Also, on a purely phenomenological basis,
the $\tfrac{5}{3}$-L\'evy distribution has been reported before \cite{CiDe05,DeDe07,ZaDe13}, although at a closer look not exactly the same quantity as here is monitored. Here we focus on quantities predicted by our theory and, in addition, implement several numerical innovations.\\
\indent (i) We average over $10^7$ initial conditions which are drawn from the exact equilibrium distribution. There is no equilibration time step.\\
\indent (ii) We employ the field theory version of anharmonic chains and measure the locally conserved fields in this representation.
We use normal mode coordinates, so to unambiguously separate the three peaks, and restrict our simulation time up to the first collision between the sound peaks.\\
\indent (iii) It is easier to fit theoretical predictions than to measure accurately scaling exponents. In our context, one has six correlation functions depending on space-time. Nonlinear fluctuating hydrodynamics suggests to use the finest 
resolution for either spatial lattice or Fourier modes
and only a few time points. In other MD simulations the converse is pursued, namely fine time, respectively frequency,
resolution and only a few smallest wave numbers. With such data the peak structure is well resolved in frequency space,
but the translation back to $(x,t)$-space cannot be  readily achieved.  

While writing, there are further MD simulations on the way. H.~van Beijeren and H.~Posch proposed the shoulder potential
for which they run extensive MD simulations. Accurate scaling plots are reported, but the non-universal coefficients
still deviate from their theoretical value  \cite{vB13}. A.~Dhar et.~al.\ \cite{Dh14} simulate FPU chains with 8192 particles and up to $t = 1600$. The interaction potential
is of the form $V(x) = \tfrac{1}{2} x^2 + \tfrac{1}{3}\mathsf{a} x^3 + \tfrac{1}{4}x^4$. The simulated parameter sets include the asymmetric case $\mathsf{a} = 2$, $p = 1$, $\beta = 2$ and the case of even potential at zero pressure, $\mathsf{a} = 0$, $\beta = 1$, $p = 0$.
Such simulations are challenging, since one has to solve the differential equations of motion. The results indicate that the heat peak scales as the $\tfrac{5}{3}$-L\'evy distribution with wave-like small perturbations receding outwards. On the other hand the sound peaks are still slightly asymmetric. The decay outside the sound cone is well approximated by $f_{\mathrm{KPZ}}$, while the opposite shoulder still exhibits slow tails resulting from the interaction with the heat peak.  S.~Lepri \cite{Le13}
simulates the FPU chain with $\mathsf{a} = 2$, $\ell = 1$, and $\mathsf{e} = 0.1$, corresponding to the pressure
$p = -0.0077$, alternatively $\mathsf{e} = 0.5$, corresponding to the pressure
$p = -0.026$. Lepri works in frequency and wave number space. The sound peak is highly resolved. For the lowest wave numbers, the scaling plot fits well with the Fourier transformed KPZ scaling function. However the measured non-universal
$\lambda_\mathrm{s}$ is off by a  factor of roughly 3 in each of the two cases. More detailed results are reported in \cite{Str14}. Motivated by quantum fluids,
M.~Kulkarni and A.~Lamacraft simulate the nonlinear Schr\"{o}dinger equation on a lattice.   Only the two sound peaks
are observed and an effective hydrodynamic model works with number and momentum as only conserved fields.
They report on the sound peak in frequency space for a few lowest wave numbers \cite{KuLa13}, in spirit similar to
\cite{Le13}. A fit to the corresponding KPZ scaling function turns out to be fairly precise \cite{KuSp14}.

An interesting variant is studied by G.~Stoltz
based on \cite{BeSt12}. His random field is specified by
$\{y_j,j\in\mathbb{Z}\}$ with $y_j \in \mathbb{R}$. The deterministic part of the evolution is governed by
\begin{equation}\label{7.7}
\frac{d}{dt} y_j = V'(y_{j+1}) - V'(y_{j-1})\,.
 \end{equation}
In addition there are random exchanges $y_j,y_{j+1}$ to $y_{j+1},y_j$ independently at each bond with rate 1. The conserved fields are $y_j$ and $V(y_j)$. The dynamics is non-reversible. The invariant measures are identical to the $\{r_j\}$-part 
of the anharmonic chain. The canonical parameters are $p,\beta$, as before, conjugate to the stretch
$\ell$ and internal potential energy $\mathsf{e}$, 
\begin{equation}\label{7.8}
\ell = \langle y_j\rangle_{p,\beta}\,,\quad\mathsf{e} = \langle V(y_j) \rangle_{p,\beta}\,.
 \end{equation}
There are no momenta. The Euler equations have only two components and read
 \begin{equation}\label{7.9}
\partial_t \ell + 2\partial_x p =0\,,\quad \partial_t \mathsf{e} - \partial_x p^2 =0
 \end{equation}
with $p = p(\ell,\mathsf{e})$.
Following the standard route one obtains the mode velocities $c_1 =0$, $c_2 = 2(-p\partial_\mathsf{e} p +
\partial_\ell p)$ and the $G$-couplings $G^1_{11} =0$, $G^1_{12} =0$, $G^1_{22} \neq 0$, while $G^2_{\alpha\alpha'}$
is generically different from 0. Thus the peak with label 2 is expected to have KPZ scaling, in analogy to our sound peaks.
The peak with label 1 will be $\tfrac{5}{3}$-L\'evy. However, since there is no symmetrically located  third peak, it will be the asymmetric 
$\tfrac{5}{3}$-L\'evy distribution at maximally allowed asymmetry, see Appendix D in \cite{Sp13}. MD simulations 
using the exponential potential $V_{\mathrm{exp}}(x) = \mathrm{e}^{-x} +x$ confirm such predictions \cite{St14}.

If $V_{\mathrm{exp}}$ is replaced by the harmonic potential $V_{\mathrm{ha}}(x) = x^2$, then one switches to a different universality class, which is the two mode version of our square-well potential at zero pressure and $\ell = a/2$. Nonlinear fluctuating hydrodynamics  predicts a diffusive peak and a $\frac{3}{2}$-L\'evy peak at maximal asymmetry. Mathematically rigorous proofs
have been posted recently  \cite{BGJ14,JKO14}.

Nonlinear fluctuating hydrodynamics is fairly insensitive to the underlying dynamics and only relies on having 
uniform, current carrying steady states. Thus it applies to quantum fluids, but also to nonreversible stochastic particle systems, as
lattice gases with several locally conserved components. The latter systems are accessible through Monte-Carlo simulations.
In \cite{FeSS13} the AHR model  \cite{AHR99} is studied.
The steady state is computed
via matrix product ansatz. Hence all coefficients are known analytically. For the normal modes one finds $c_1 \neq c_2$ and also $G^{1}_{11} \neq 0$, $G^{2}_{22} \neq 0$. However the subleading coefficients vanish,
$G^{1}_{22} = 0 = G^{2}_{11}$. In Monte Carlo simulations one observes a rapid relaxation to $f_{\mathrm{KPZ}}$ for each mode. The non-universal $\lambda$ coefficient is relaxed and attains precisely the value as deduced from the theory.
A coupled two-lane TASEP is studied in \cite{PSS14}, which besides two KPZ peaks allows one to also realize the cases of  a KPZ peak with a $\tfrac{5}{3}$-L\'evy peak and, more exotically, of a KPZ peak with a diffusive peak.

\paragraph{Acknowledgments.} We are grateful to H.~van Beijeren and H.~Posch for sharing with us their insights on simulating a fluid with hard-shoulder potential. We greatly profited from discussions with S.~Denisov, A.~Dhar, P.~Ferrari, D.~Huse, J.~Krug, M.~Kulkarni, S.~Lepri, R. Livi,
A.~Politi, T.~Sasamoto, G.~Sch\"{u}tz, and G.~Stoltz. Computing resources of the Leibniz-Rechenzentrum are thankfully acknowledged.


\appendix

\section{Square-well potential}
\label{secA}

The square-well potential serves as an example, for which the transformation to normal modes and the second order expansion 
are still reasonably explicit.

We choose the spatial unit such that $a = 1$. Then the thermodynamic potentials are given by 
\begin{eqnarray}\label{A.1}
&&\hspace{-10pt} p(\ell,\mathsf{e}) = 2 \mathsf{e} h(\ell)\,,\quad 2\mathsf{e}\beta = 1\,,\nonumber\\
&&\hspace{-10pt} \partial_\ell p = 2 \mathsf{e} h'\,,\hspace{3pt}  \partial_\mathsf{e} = 2h\,,\quad \partial_\ell^2 p = 2 \mathsf{e} h'{'}\,,\hspace{3pt}
\partial_\ell\partial_\mathsf{e} p = 2h'\,,\hspace{3pt}\partial_\mathsf{e} ^2p = 0\,.
\end{eqnarray}
For the function $h$ we use only that $h' <0$. The concrete $h$ is given below \eqref{2.25}. For the hard-point gas,
$h = 1/\ell$. The sound speed is
\begin{equation}\label{A.2}
c_{\bar{m}} = c / \sqrt{\bar{m}}\,,\quad c^2 = 2\mathsf{e}(-h' +2 h^2)\,.
 \end{equation}
The susceptibility is
\begin{equation}\label{A.3}
C=
\begin{pmatrix} (-h')^{-1} & 0 & 0 \\
               0 & 2\mathsf{e} \bar{m}& 0 \\
                 0 & 0 & 2\mathsf{e}^2
\end{pmatrix}\,,
\end{equation}
 and the linearized Euler equations are governed by
\begin{equation}\label{A.4}
A=
\begin{pmatrix} 0 & -\bar{m}^{-1} & 0 \\
               2 \mathsf{e} h'& 0 & 2h \\
                0 & 2 \mathsf{e}h \bar{m}^{-1}& 0
\end{pmatrix}\,.
\end{equation}
From the eigenvectors of $A$ one obtains the transformation matrix $R$ as
\begin{equation}\label{A.5}
R = 
\frac{1}{2c\sqrt{\mathsf{e}}}\begin{pmatrix}
2\mathsf{e}h'&- c_{\bar{m}}&2h\\
4 \mathsf{e}h \sqrt{-h'}&0& 2 \sqrt{-h'} \\
2\mathsf{e}h'&c_{\bar{m}} &2h\\
\end{pmatrix}\,,
\end{equation}
\begin{equation}\label{A.6}
R^{-1}= \frac{\sqrt{\mathsf{e}}}{c}
\begin{pmatrix}
-1&2h/\sqrt{-h'}&-1\\
- c_{\bar{m}} \bar{m}&0& c_{\bar{m}} \bar{m}\\
2\mathsf{e}h&2\mathsf{e}\sqrt{-h'}&2\mathsf{e}h\\
\end{pmatrix}\,.
\end{equation}

Next we compute the $G$ matrices for the nonlinear coupling constants. Firstly, by direct differentiation of $p$,
\begin{equation}
\label{A.7}
H^\ell =0\,,\quad H^\mathsf{u}= 
2\begin{pmatrix}  \mathsf{e} h'{'}& 0 & h'\\
               0 & -h\bar{m}^{-1} & 0 \\
                h' & 0 & 0
\end{pmatrix}
\,,\quad H^\mathsf{e}= \frac{2}{\bar{m}}
\begin{pmatrix} 0 &\mathsf{e}h' & 0 \\
               \mathsf{e}h'& 0 & h \\
                0 & h & 0
\end{pmatrix}
\end{equation}
and transformed as
\begin{equation}
\label{A.8}
(R^{-1})^{\mathrm{T}}H^\mathsf{u} R^{-1}= \frac{2\mathsf{e}^2}{c^2} 
\begin{pmatrix} a_3 & a_1& a_4 \\
              a_1 & a_2 & a_1 \\
                a_4 & a_1 & a_3
\end{pmatrix},\quad
(R^{-1})^{\mathrm{T}}H^\mathsf{e} R^{-1} = 2\mathsf{e}c\bar{m}^{-1/2}
\begin{pmatrix} -1 & 0& 0 \\
              0 & 0 & 0 \\
                0 & 0 & 1
\end{pmatrix}\,,
\end{equation}
where
\begin{equation}
\label{A.9}
\begin{split}
a_1 &= 2(-h')^{-1/2}\big(-hh'{'} + h'^2 +2 h^2h'\big)\,,\quad a_2 = 4(-h')^{-1}\big( h^2h'{'} - 2h h'^2\big)\,,\\
a_3 &= h'{'} - 2 h h' -4h^3\,, \qquad a_4 = h'{'} -6hh' +4h^3\,.
\end{split}
\end{equation}
One still has to apply $R$ to $(R^{-1})^{\mathrm{T}}\vec{H} R^{-1}$. The final result reads
\begin{align}
\label{A.10}
G^0 &= \sqrt{-h'\,\mathsf{e}/\bar{m}} 
\begin{pmatrix} -1 & 0& 0 \\
              0 & 0 & 0 \\
                0 & 0 & 1
\end{pmatrix}\,,\\
G^\sigma &= \frac{1}{2} \sqrt{\mathsf{e}/\bar{m}} \left(\sigma\frac{1}{2(-h' + 2h^2)}
\begin{pmatrix} a_3 & a_1& a_4 \\
              a_1 & a_2 & a_1 \\
                a_4 & a_1 & a_3
\end{pmatrix}
+2h
\begin{pmatrix} -1 & 0& 0 \\
              0 & 0 & 0 \\
                0 & 0 & 1
\end{pmatrix} 
\right)\,.
\end{align}

At $2\ell = a$, $p = 0$ the sound speed simplifies to $c_{\bar{m}} = 2 \sqrt{6\,\mathsf{e}/\bar{m}}$ and the coupling matrices to
\begin{equation}
\label{A.11}
G^{0} = 2 \sqrt{3\,\mathsf{e}/\bar{m}}
\begin{pmatrix}
 -1 & 0 & 0 \\
 0 & 0 & 0 \\
 0 & 0 & 1 \\
\end{pmatrix}, \quad
G^{\sigma} = \sigma \sqrt{3\,\mathsf{e}/\bar{m}}
\begin{pmatrix}
 0 & 1 & 0 \\
 1 & 0 & 1 \\
 0 & 1 & 0 \\
\end{pmatrix}\,.
\end{equation}
For the hard-point gas, $a = \infty$, hence $p(\ell) = 1/\ell$, and one finds $a_1 = -6\ell^{-3}$, $a_2 = 0$, $a_3 =0$, $a_4 = 12\ell^{-3}$.

\section{Speed of sound, $R$ matrix, and $G$ couplings}
\label{secB}

For each model, at our parameters, we record $c$, $R$, and $G$. One has the relation $G^{-1} = - (G^{1})^{\mathcal{T}}$,
where ${}^\mathcal{T}$ stands for transpose relative to the anti-diagonal.
Thus only $G^1$ is listed. The entries are rounded to four digits for visual clarity.

\paragraph{Shoulder potential.} Our parameters $p = 1.2$, $\beta = 2$ imply $ c \simeq 1.743$ and
\begin{equation}
R =
\begin{pmatrix}
 -0.8067 & -1 & 0.7800 \\
 2.1031 & 0 & 1.7526 \\
 -0.8067 & 1 & 0.7800 \\
\end{pmatrix}\,, \quad
R^{-1} =
\begin{pmatrix}
 -0.2869 & 0.2554 & -0.2869 \\
 -0.5 & 0 & 0.5 \\
 0.3443 & 0.2641 & 0.3443 \\
\end{pmatrix}\,,
\end{equation}
as well as
\begin{equation}
G^{1} =
\begin{pmatrix}
 -0.3131 & -0.0123 & 0.3664 \\
 -0.0123 & 0.2014 & -0.0123 \\
 0.3664 & -0.0123 & 0.3664 \\
\end{pmatrix}\,,
\quad
G^{0} =
\begin{pmatrix}
 -0.7635 & 0 & 0 \\
 0 & 0 & 0 \\
 0 & 0 & 0.7635 \\
\end{pmatrix}. \\
\end{equation}

\paragraph{Hard-point gas with alternating masses.} The general expression for $R$ reads
\begin{equation}
R = \frac{1}{\sqrt{6}} \begin{pmatrix}
 -\beta p & -\sqrt{3 \beta / \bar{m}} & 2 \beta \\
 2 \beta p & 0 & 2 \beta  \\
 -\beta p & \sqrt{3 \beta / \bar{m}} & 2 \beta \\
\end{pmatrix}.
\end{equation}
The $G$ matrices only depend on the sound speed, and the general formula is
\begin{equation}
G^{1} =
\frac{c_{\bar{m}}}{2 \sqrt{6}}
\begin{pmatrix}
 -2 & -1 & 2 \\
 -1 & 0 & -1 \\
 2 & -1 & 2 \\
\end{pmatrix}\,, \quad
G^{0} =
\frac{c_{\bar{m}}}{\sqrt{6}}
\begin{pmatrix}
 -1 & 0 & 0 \\
 0 & 0 & 0 \\
 0 & 0 & 1 \\
\end{pmatrix}, \\
\end{equation}
with sound speed $c_{\bar{m}} = c / \sqrt{\bar{m}}$ and $c = \sqrt{3\beta}\,p$. Specifically for $m_0 = 1$, $m_1 = 3$, $p = 2$, $\beta = 1/2$, one obtains $c_{\bar{m}} = \sqrt{3} \simeq 1.732$ and
\begin{equation}
R =
\begin{pmatrix}
 -0.4082 & -0.3536 & 0.4082 \\
 0.8165 & 0 & 0.4082 \\
 -0.4082 & 0.3536 & 0.4082 \\
\end{pmatrix}\,, \quad
R^{-1} =
\begin{pmatrix}
 -0.4082 & 0.8165 & -0.4082 \\
 -1 & 0 & 1 \\
 0.8165 & 0.8165 & 0.8165 \\
\end{pmatrix},
\end{equation}
\begin{equation}
G^{1} =
\begin{pmatrix}
 -0.7071 & -0.3536 & 0.7071 \\
 -0.3536 & 0 & -0.3536 \\
 0.7071 & -0.3536 & 0.7071 \\
\end{pmatrix}\,, \quad
G^{0} =
\begin{pmatrix}
 -0.7071 & 0 & 0 \\
 0 & 0 & 0 \\
 0 & 0 & 0.7071 \\
\end{pmatrix}\,.
\end{equation}

\paragraph{Square-well potential, $a = 1$ and $p = 0$.} The general formula for $R$ is provided in Eq.~\eqref{A.5} and for $G$ in Eq.~\eqref{A.11}. Inserting $\beta = 2$ and alternating masses $m_0 = 1$, $m_1 = 3$ results in $c_{\bar{m}} = \sqrt{3}$ and
\begin{equation}
R =
\begin{pmatrix}
 -2.4495 & -0.7071 & 0 \\
  0 & 0 & 2.8284 \\
 -2.4495 & 0.7071 & 0 \\
\end{pmatrix}\,, \quad
R^{-1} =
\begin{pmatrix}
 -0.2041 & 0 & -0.2041 \\
 -0.7071 & 0 & 0.7071 \\
 0 & 0.3536 & 0 \\
\end{pmatrix}\,,
\end{equation}
\begin{equation}
G^{1} = 
\begin{pmatrix}
 0 & 0.6124 & 0 \\
 0.6124 & 0 & 0.6124 \\
 0 & 0.6124 & 0 \\
\end{pmatrix}, \quad
G^{0} = 
\begin{pmatrix}
 -1.2247 & 0 & 0 \\
 0 & 0 & 0 \\
 0 & 0 & 1.2247 \\
\end{pmatrix}.
\end{equation}

\section{Scaling functions}\label{secC}
\setcounter{equation}{0}
The non-universal $\lambda$ coefficients are defined relative to a conventional choice of the scaling functions,
which we list for convenience.\medskip\\
The \textit{Gaussian} of unit variance is defined as
\begin{equation}\label{c.1} 
f_\mathrm{G}(x) = (2\pi)^{-1/2}\,\mathrm{e}^{-x^2/2}\,.
\end{equation}\medskip\\
The symmetric \textit{L\'evy distribution} with index $\alpha$, $0 < \alpha <2$, is given by
\begin{equation}\label{c.2} 
f_{\mathrm{L},\alpha}(x) = \frac{1}{2\pi}\int_{\mathbb{R}}dk \mathrm{e}^{\mathrm{i}kx}\mathrm{e}^{-|k|^\alpha}\,.
\end{equation}
$f_{\mathrm{L},\alpha}(x) \simeq |x|^{-\alpha -1}$ for large $|x|$. 
\medskip\\
The \textit{KPZ scaling function} $f_{\mathrm{KPZ}}$ is tabulated in~\cite{Prhp}, denoted there by $f$. It holds
\begin{equation}
f_{\mathrm{KPZ}}\geq 0,\hspace{6pt} \int_{\mathbb{R}} dxf_{\mathrm{KPZ}}(x)=1,\hspace{6pt} f_{\mathrm{KPZ}}(x)=f_{\mathrm{KPZ}}(-x),\hspace{6pt} \int_{\mathbb{R}} dxf_{\mathrm{KPZ}}(x)x^2=0.510523\ldots\,. 
\end{equation}
$f_{\mathrm{KPZ}}$ looks like a Gaussian with suppressed tails, more precisely a large $|x|$ decay as $\exp[-0.295|x|^{3}]$~\cite{PrSp04}. The natural definition of $f_{\mathrm{KPZ}}$ involves the Fredholm determinant of the Airy kernel,  which then implies  our particular value of the variance.


\end{document}